\documentclass[pre,superscriptaddress,twocolumn,showkeys,showpacs]{revtex4}
\usepackage{epsfig}
\usepackage{latexsym}
\usepackage{amsmath}
\def\go{\rightarrow}
    \def\e{\epsilon }
\def\g{\gamma } \def\G{\Gamma}   
    
\def\s{\sigma } 

\def\bs{\backslash}
\def\sN{{\cal N}}
\def\s1{{\textbf{1}}}
\newcommand{\bea}{\begin{eqnarray*}}
\newcommand{\eea}{\end{eqnarray*}}
\newcommand{\benum}{\begin{enumerate}}
\newcommand{\eenum}{\end{enumerate}}

\begin {document}
\title {Noise in Naming Games, partial synchronization and community detection in social networks}
\author{Weituo Zhang}
\affiliation{Mathematical Sciences, RPI, 110 8th St., Troy, NY, 12180}
\author{Chjan C. Lim}
\affiliation{Mathematical Sciences, RPI, 110 8th St., Troy, NY, 12180}
%%%%%%%%%%%%%%%%%%%%%%%%%%%%%%%%%%%%%%%%%%%%%%%%%%%%%%%%%%%%%%%%%%%%%%%%%%%%%
\begin {abstract}
The Naming Games (NG) are agent-based models for agreement dynamics,
peer pressure and herding in social networks, and protocol selection
in autonomous ad-hoc sensor networks. By introducing a small noise
term to the NG, the resulting Markov Chain model called Noisy Naming
Games (NNG) are ergodic, in which all partial consensus states are
recurrent. By using
Gibbs-Markov equivalence we show how to get the NNG's
stationary distribution in terms of the local specification of a
related Markov Random Field (MRF). By ordering the partially-synchronized states according to
their Gibbs energy, taken here to be a good measure of social
tension, this method offers an enhanced method for
community-detection in social interaction data.
We show how the lowest Gibbs energy multi-name states separate and
display the hidden community structures within a social network.
\end{abstract}
\maketitle
\section{Introduction}

The possibility of using the Naming Game
(NG)\cite{Baronchelli2006-2,Dall'Asta2006}, a family of agent-based
models, for modeling agreement dynamics, leader election and the
effects of peer pressure in social networks
\cite{Angluin1980,Castellano2005,Ben-Naim1996},
protocol and security key selection in encrypted communication in
autonomous sensor networks\cite{Collier2004}, vocabulary
selection in linguistics \cite{Nowak2001,Matsen2004} and
the role of herding in market crashes in financial networks
%his new book Didier Sornette, Why stock Markets crash - Critical events in Financial Complex Systems]
have recently attracted great interest in the scientific community.
Formation of opinions and agreement dynamics in complex financial
and social networks are relevant examples of collective /
cooperative behaviors that are driven by the individual's innate
propensity to imitate and hence herd in the absence of reliable
information or when there is information asymmetry. Finite-size
effects and the underlying network's topology such as scale-free
\cite{AlbertJeongBarabasiNature} and small world
\cite{WattsStrogatzNature}, have significant impact on the expected
time to consensus (single-name or total-synchrony state) and other
related critical exponents. Indeed simulations  showed departure
from the values obtained by classical scaling arguments based on the
emergence of the giant cluster in percolation theory
\cite{Meester1996,Penrose2003,Barabasi} and on the dynamical theory
of coarsening in high-q Pott's models
\cite{Stanley,Kaski1985,Kumar1987,Roland1990,Lu2008}, highlighting
the need for more simulation studies and also Markov Chain analysis
\cite{Bremaud}.

A recent application of the Naming Games is to find
hidden communities in social networks \cite{Lucomm,Lu2008},
\cite{Girwan2002}. The main objectives of this paper are to sharpen the community detection capability of the original Naming Games by the explicit and purposeful addition of rare noise events, and to analyze rigorously the consequences of adding this noise.
Community structure is extremely difficult to define
and uncover and remains one of the outstanding open problems in network science.
Amongst existing methods, the most powerful algorithms are
based on the notion of modularity \cite{NewmanPNAS2006} which provides an ordering of
the closeness between the given network's community structure and a large family of motifs or partitions.
The optimization problem involved in finding the motif with the largest modularity grows rapidly with
the size of the network. While several working heuristic methods have been proposed, they
are not better than a spectral method based on the eigenvectors of a Laplacian-like matrix \cite{NewmanPNAS2006}. This procedure of splitting the network into two parts is repeated in a binary tree way until the subgraphs are indivisible.

In particular, direct simulated-annealing of the modularity optimization problem will be computationally expensive, just as other Monte-Carlo methods applied to the original Naming Games have yielded good results only after costly long simulations \cite{Lu2009}. In this article we indicate how Monte-Carlo methods and other importance sampling algorithms can be numerically efficient on the Noisy Naming Games but not on the original Naming Games. Comparing the Noisy Naming Game method to the modularity optimization procedure, it is worthwhile noting that these two methods start from opposite ends of the community structure: the modularity - binary tree method finds first, the optimal division of the whole network into two parts but the Noisy Naming Games, beginning from random initial sublists of the allowed words, first finds the (last) optimal division into small communities. The modularity increases as the binary tree method proceeds towards the indivisible units but the Gibbs potential decreases as the Noisy Naming Game proceeds towards the low-lying states near to its ground state of total consensus. In this way, the Noisy Naming Games are complementary to the modularity based spectral binary tree method. If one is looking for the finer divisions of the network into small communities then the Noisy Naming Games will find them provably faster than other methods including the modularity-based methods.

But before the Naming Games can provide an applicable procedure for community detection, three problems need to be addressed. The first is the transient nature of its multi-name partial
consensus states that correspond to possible community structures.
%will be discussed next in relation to the general
%issue of dynamic synchrony.
The second concerns the existence and computability of a stationary measure that can be used -- like the notion of modularity -- to rank the community test-patterns in the form of multi-name partial consensus states.
The third issue concerns the computational efficiency of the Naming Games viewed herein as an inverse method for finding communities in networks.
It turns out that these problems can be overcome by adding rare noise events to the standard stochastic framework provided by the Markov chain model of the original Naming games. Unlike the original Naming Games which have been compared to q-Potts models \cite{Stanley}, the Noisy version introduced here is more closely related to spin-glass models with an additional layer of randomness \cite{Nishimoribook}. As is well-known certain aspects of information processing and combinatorial optimization (such as the graph-partitioning problem that is closely related to community detection) can be addressed from the vantage point of the spin-glass framework. In many instances, the ground-state of the spin-glass model provides information for the optimization problem. But computing or finding the ground-state is often an NP-hard problem, of an equivalent degree of difficulty as the optimization problem itself. The spin-glass framework offers the alternative path --- often seen to work well in explicit positive usage of noise to solve hard deterministic problems -- of exploiting not only the inherent spin entropy of the model (when the interactions are quenched) but also the additional randomness and corresponding entropy arising from the ensemble of quenched interactions for which the replica method \cite{Nishimoribook} is designed to average over.
We exploit these relationships in discussing each of these three issues in formulating an efficient procedure based on the Noisy Naming Games for finding community structures.

First, because the total consensus states of the original Naming Games are the only
absorbing states, the detection of communities or partial consensus states is complicated
by their transient nature. Meta-stable states representing multi-names or partial
consensus configurations arise in the original Naming Games on networks with strong
community-structure but have been shown numerically to evolve into
other partial-synchrony states after an initial phase
\cite{Lucomm,Lu2008}, according to a time hierarchy in general.
Complete agreement on leader selection or total consensus occurs on
time-scales that are relatively short after a long meta-stable
stage, with negative consequences on the design of algorithms for
detection of hidden communities, for example, in social networks
with Small World topologies \cite{StanleyPNAS2000},
\cite{Girwan2002}. Related results on the synchronization of
dynamical networks and discrete time maps on networks that highlight
the roles of the network topology and the coupling weights have been
discovered \cite{Motter06}, \cite{Earn2006}. In other significant
contexts such as the modeling of brain functions by coupled
oscillators on a network, the transient nature of partially
synchronous states play significant positive roles.

The first issue that has to be solved is to find a version
of the Naming Games where all partial consensus states are recurrent. This can be done if we come up with a
version of the Naming Games which is an ergodic Markov chain. A nice property of ergodic Markov chains is the existence and uniqueness of an invariant measure with positive weights on all the recurrent states. Then the next issue in designing an efficient procedure for community detection starting from the original Naming Games is finding a closed-form expression for this invariant measure and proving that given the network topology (such as neighborhood structure) and a reasonably small set of allowed words in the original game, this invariant measure ranks the partial consensus states in an averaged sense (macrostates) incorporating all entropic contributions in the Noisy Naming Games. One key property of this invariant measure is the free energy gaps between the low-lying states near the ground-state -- larger gaps lead to better resolution of the community structure. We show by exact calculations on small cliques that the invariant measure of the Noisy Naming Game inherits this property from its closed-form Gibbs potential. Although the large-scale implementation of the Monte-Carlo simulation of the Noisy Naming Game on a real world social network remains to be done, our small-scale testing of this method on computer-generated randomized networks of about 60 nodes allow us to infer that the method is computationally efficient.
In the context herein, rigorous results stating new
conditions for the existence and rank-ordering of recurrent
multi-names or partial-consensus states in the Naming Games are
useful in view of potential applications to detecting hidden
communities in interaction data such as Twitter and Facebook.

In this article, we show that a class of arbitrarily small noisy
perturbations of the Naming Games, called Noisy Naming Games (NNG),
satisfies these conditions. The main results here are analytical
ones: First, the Noisy Naming Games differing from the NG by
arbitrarily small stochastic perturbation, has a unique invariant
distribution with positive weights for partial consensus states.
Second, we construct the NNG's invariant distribution as the Gibbs
potential of a related Markov Random Field (NNGGS) called the Noisy
Naming Game with saturated training from its local specification.
Third, we show the Gibbs energy (which is taken to be a reasonably
efficient measure of social tension in this article) provides an
ordering of the recurrent multi-name or partial-consensus states of
the NNG whereby the lowest energy and thus most probable ones, also
have single-name or single-color cliques. This last result has
significant impact on the detection of (hidden) community structure
in a social network by Monte-Carlo simulations. After the initial
equilibration process, the fraction of time spent by the Monte-Carlo
simulator in any given partial-consensus state is proportional to
the probability it is assigned by the Gibbs invariant measure. Thus,
besides identifying the low-lying Gibbs energy states which are the
most indicative of underlying community structure - near-cliques or
tightly connected clusters have a common single-name or color -- the
simulation when run long enough will also provide an ordering of these
states. We will show with explicit calculations on simple examples, that the exact Gibbs energy of the
multi-name states not only orders them but have the essential property of significant energy gaps
between the low-lying states.
%Simple examples of clique potentials will be discuss
%in relation to community detection in social networks.

In brief, the introduction of additional randomness in the form of
rare noisy events to the original Naming Games provides the key to
successfully enhancing community-detection through the existence and
rank-ordering of a large family of stable partial-consensus states
where tightly-connected clusters in the network show up as
single-named or single-colored subgraphs in the social network.
Moreover, these partial consensus states can be found and ranked by simulating
the Noisy Naming Games using an equivalent Gibbs sampler method which was used to produce the figures in the last section.
In contrast to the generic need
to avoid or prevent noise in most technological systems, the precise
and explicit use of noise to achieve a positive aim in this project
has few precedents. Significant examples are the classical Parrondo
games and the Brownian Ratchet \cite{LeeAllisonAbbottStanley}.
%Third, we show that this Gibbs Sampler\cite{Casella1992} algorithm
%for simulating the NNG's distribution only requires $O(2^{cM}M)$
%calculations initially of the local specification which are reused
%in each subsequent sweep of the NNGGS. Here, $c$ is the size of the
%finite set of allowed words, Name, in the NNG and $M$ is the maximum
%degree of the network graph, thence the significant computational
%efficiency of the Gibbs Sampler for normally sparse ($M<<N$) social
%networks with $N$ sites, relative to the NNG itself.
% -- part of the computational economy
%realized here is related to that in many divide-and-compute
%algorithms such as the FFT.

%\benum
%\item

\section{ Markov chain model}
We construct a Markov chain model for the
NG \cite{Baronchelli2006-2} where a transition consists of the
change of local state at only one site in the network and the set
Name of all allowed words is fixed at the outset. It is easier to
analyze by the Gibbs Sampler method than the usual NG to which it is
equivalent. Consider a network based on a connected graph containing
N sites $S=\{s_i\}$.
%is a finite set of all sites in the network.
The
neighborhood structure of the network is given by
$\{\sN_{s_i}\}_{s_i\in S}$. Each site $s_i$ will have a word list
chosen from the finite set $Name=\{A,B,C...\}$. Set $\G=\{\g_k\}$, which
consists of all non-empty subsets $\g_k$ of $Name$, represents all
possible "word lists" of a site. The configuration function
$X(s_i): S\rightarrow \G$ mapping each site to its word list gives
the local state of $s_i$. A configuration restricted to a subset of sites $\Sigma\subset S$
is denoted by $X(\Sigma)$. Therefore the network state $G$- a word list assigned to each site- is given
by the configuration $G=\{X(s_i)\}_{s_i\in
S}=X(S)$. In the Naming Game (NG), we change the state of the network $G$ as follows:
\benum \item In this step we randomly choose a site $s_i\in S$ as a
"listener", with probability $q(s_i)$.
\item Next, choose a site
$s_j\in \sN_{s_i}$ with equal probability as a "speaker". The
"speaker" will randomly choose a word $W$ from its word list with
equal probability and send it to the "listener". The latter will
change its state $X(s_i)$ into $X'(s_i)$ by the following way:
$$X'(s_i)=\begin{cases} X(s_i)\cup \{W\} & W\notin X(s_i)\\
 \{W\} & W\in X(s_i)
\end{cases}$$
\eenum We call step (2) the local transition step and the process
(1),(2) together comprises a transition between
$G=\{X(s_i),X(s_j)\}_{s_j\in S\bs\{s_i\}}$ and its neighboring state
$G'=\{X'(s_i),X(s_j)\}_{s_j\in S\bs\{s_i\}}$. The global transition
probability $P(G,G')$ of the NG depends on the probability for
choosing $s_i$ and the local transition probability from state
$X_{0}(s_i)$  to $X'{(s_i)}$ under neighborhood configuration
${X_{0}(\sN_{s_i})}$ is $P(G, G')= q(s_i) P(X_{0}(s_i),X'(s_i)|
{X_{0}(\sN_{s_i})} )$. \\
By randomly choosing an initial state
$G_0\in \G^S$ and applying steps (1) and (2) in each time period, we
obtain a homogeneous Markov chain of the Naming Game $\{G_0, G_1...
G_n...\}$, where the transition matrix from $G_n$ to $G_{n+1}$ is
given by the formula above.

%\item
Next, we show that any invariant measure of this Markov chain is a
linear combination of $\s1_{G_{W}}$ where $\s1$
is the indicator function and $G_W$ are the "single-name states" in
which every site has only $W$ in its word list. Let $G_{\g_k}$ for
$\g_k \in \G$ be the set of network states that satisfies the
condition $\bigcup_{s_i\in S}X(s_i)=\g_k .$ It is clear from the NG
Markov chain above that single-name states are the only absorbing
ones.\\
%is that correct?
To proceed, every network state $G\in G_{\g_k}$ with $|\g_k|\geq2$
is accessible to at least one absorbing state $G_W$, i.e. it will
have a path $\{G,G_1, G_2...G_n,G_W\}$ with nonzero probability
$p=(G,G_1)P(G_1,G_2)...P(G_n,G_W)$. Since this probability $p$ should be counted in the probability of leaving the
state $G$ and never returning to it, any state $G$ which is not
absorbing is transient. In an invariant measure of a Markov chain, the weight of any transient state is zero; thus,
 the only states with positive weights in an invariant measure of the NG Markov chain
are the single-name states $G_W (W\in Name)$.
Since $\s1_{G_{W}}$ itself is always an
invariant measure, any
invariant measure of the Naming Game is a linear combination of
$\s1_{G_{W}}$ over all $W\in Name$. In other words, the Naming Game
Markov Chain, starting from any initial state will
eventually go to a single-name state.

\section{ Noisy Naming Games (NNG)}
In view of the above Markov chain analysis, we need to perturb the original Naming Game to enhance its community-detection capabilities. It turns
out that the class Noisy Naming Game (NNG) of arbitrarily small random perturbations of NG, given next, has the required key property of persistent
multi-name steady states.
In each step in the NNG Markov chain, the listener has a very small
probability $\e$ to receive a noise word rather than the speaker's
message where the noise word is a random word chosen uniformly from $Name$,
and a probability of $1 - \e$ of proceeding as in the original NG.
Then given two arbitrary network states, one can construct a path
from one to the other with nonzero probability by forcing the
network to receive a sequence of noise words as required. In this
way, the NNG Markov chain is forced to be communicative and therefore
ergodic, by arbitrarily small $\e$. On the other hand, since $\e$ is
arbitrarily small, the event of receiving a noise word rarely happens.
So the noise will essentially change the
long term behavior but can be ignored in finite time.
From the NNG Markov chain's ergodicity, its invariant measure is positive and
unique up to a scalar multiple; hence the first main result:

$\textbf{Result}:$ The NNG has a unique invariant probability distribution
$\pi(G)>0$ directly related to its global transition probability $P(G,G').$
Moreover, each multi-name state $G_m$ is recurrent and has positive weight, i.e.,
$\pi(G_m)>0.$
%which will be related below
%to a family of local conditional distributions $f_i(X|X(\sN(s_i)), X\in \G$ over all sites $s_i \in S$.

In both the original NG and the NNG, the local transition step (2) is given by
%by a local Markov chain
$X_{0}(s_i)\rightarrow X'(s_i)$ with corresponding local transition
probability $P(X_{0}(s_i),X'(s_i)| {X_{0}(\sN_{s_i})}
)\s1_{X(\sN_{s_i})=X_{0}(\sN_{s_i})}$. By repeating the
NNG's local transition step (2) many times at fixed site $s_i \in S$
with the same neighborhood state $ X_{0}(\sN_{s_i})$, we generate a
sequence of local states $\{X_{0}(s_i),X'(s_i),...,X^{(n)}(s_i)\}$,
such that the marginal probability distribution of $X^{(n)}(s_i)$
converges to a l    ocal conditional probability distribution
$f_i(X|X_{0}(\sN(s_i))$ as n goes to infinity. Thus,
$f_i(X|X_{0}(\sN(s_i))$ over all $s_i \in S$ is a well-defined
limiting distribution of a local Markov chain with one site and
fixed neighborhood state.
%which is calculated once initially and reused in all subsequent sweeps
%We emphasize the point that nothing beyond the existence of $f_i(X|X(\sN(s_i))$ is assumed or known.

To calculate the NNG's invariant distribution,
we first construct a related Markov chain (NNGGS) verified to
be a Gibbs Sampler for a Markov Random Field \cite{Bremaud}. Keeping
step (1) the same, the NNGGS replaces the local transition step (2)
in the NNG
by a "training step":\\
\emph{{(2')  \ $X_{0}(s_i)\rightarrow X^*(s_i)$ which is the value of a random variable distributed according to the local conditional distribution $f_i(X^*|X_{0}(\sN(s_i)).$}}\\
The local training step (2') is thus equivalent to repeating the
NGG's local transition step (2) infinite number of times at fixed
site with the same neighborhood state. Its global transition
probability is given by
$$P^{'}(G,G')=q(s_i)f_i(X^{'}(s_i)|X_{0}(\sN(s_i))\s1_{X(\sN_{s_i})=X_{0}(\sN_{s_i})}$$
Since only the local state at a single random site is changed
according to a local specification, the NNGGS is indeed a Gibbs
Sampler on a Markov Random Field (MRF) with local specification
given by the family $f_i(X|X(\sN(s_i))$ for $s_i \in S$
\cite{Bremaud}. Here, the NNGGS is more an indirect analytical
method to derive the stationary distribution of the NNG in closed
form than a method for numerical simulation of the NNG, for which
there are better Markov chain Monte-Carlo methods \cite{Limnebus}.

\section{ Stationary distribution of the Gibbs Sampler} Using the
Gibbs-Markov equivalence \cite{Bremaud}, we construct in terms of
the NNGGS's local specification, a Gibbs potential
%(invariant measure of the Markov chain)
which is the stationary distribution $\pi^{'}(G)$ of the NNGGS. In
other words, under $P^{'}(G,G')$, the NNGGS generates a Markov chain
of realized states and a distribution, that converge to
$\pi^{'}(G).$ First, we need to assign a 0 local state for each
site. Then 0(E) for $E\subset S$ denotes the 0 configuration state
where all sites in E are at the 0 local state. The choice of the 0
local state is not unique-- for convenience we choose the whole
word-list $\g=Name$ as the 0 local state. For a clique $L\subset S$
of the network graph, let $x(L)$ be a certain configuration on $L$.
Then the clique potential of this configuration
is\cite{Grimmett1973}:
$$V(x(L))=-\sum_{E\subset L} (-1)^{|L-E|}
\ln[F(x(E)|0(\sN(E)))].$$
%$\ln[\pi(x(E),0(\sN(E))) (\pi(0(S)))^{-1}]$$
Here the summation is over al, subsets of the clique $L$, the
neighborhood of a subset $E=\{s_i | i=1,...,M\}$ is defined by $\sN(E)=\bigcup_{s_i\subset
E} \sN(s_i)\ \backslash E $ and $F(x(E)|0(\sN(E)))$ is given by:
$$\prod_{i=1}^M {f_i(x(s_i)|X(s_1,...,s_{i-1}),0(s_{i+1},...,s_M),0(\sN(E)))\over f_i(0(s_i)|X(s_1,...,s_{i-1}),0(s_{i+1},...,s_M),0(\sN(E)))}$$
%\ln[\pi^{'}(x(E),0(\sN(E))) (\pi^{'}(0(S)))^{-1}]$$
%where $pi^{'}(.)$ is the stationary distribution of the Gibbs Sampler NNGGS which

%$P^{'}(G,G')$ of the NNGGS.
%is proportional to $\sum_{s_i \in E}
%\ln[f_i(x(s_i)|0(\sN(s_i)))].$
Then the Gibbs potential is given by the sum (over all cliques in
the network graph) of clique potentials: $$H(x)=\sum_{L\subset S
}V(x(L)).$$
The stationary distribution of the NNGGS is thus:
$$\pi^{'}(x)=Z^{-1} \exp(-H(x)).$$ where $Z=\sum_{G} \exp(-H(G)))$ is
the Gibbs partition function\cite{Bremaud}. It remains to show that
$\pi^{'}$ is invariant under the NNG, that is, $\pi^{'}=\pi$.

\section{ Invariant distribution of the NNG} By the existence and
uniqueness of the invariant distribution of the NNG, any
distribution that is invariant under the NNG must be equal to
$\pi(G).$ Here, we verify that the Gibbs distribution $\pi^{'}(G)$
of the NNGGS is invariant under the NNG, thence the explicit
construction of $\pi(G)$ as the second main result. We will do this
by showing and using detailed balance several times. Consider a
chosen network state $G_0=\{X_0(s_1),...,X_0(s_i),...X_0(s_N)\}$ and
any neighboring network state
$G_{ik}=\{X(s_1),...,X(s_i),...X(s_N)\}$ where all but site $s_i$
have the same local states as $G_0$ and the differing local state is
the word list $\g_k \in \G$. By detailed balance of the NNGGS:
$$\pi'(G_0)q(s_i)f_i(\g_k|X_0(\sN(s_i)))
=\pi'(G_{ik})q(s_i)f_i(X_0(s_i)|X_0(\sN(s_i))).$$
In terms of the
global transition probability $P(G,G')$ of the NNG, the change of
measure under a single step of the NNG  $\Delta\pi'(G_0)$ is \bea
&&\sum_i\sum_k
\left[P(G_{ik},G_0)\pi'(G_{ik})-P(G_0,
G_{ik})\pi'(G_0)\right]\\
&=& \sum_i q(s_i)
%\ul
\sum_k \Big[P(\g_{k},X_0(s_i)|
X_0(\sN(s_i)))\pi'(G_0) \\
&&-P(X_0(s_i),\g_k|X_0(\sN(s_i)))\pi'(G_{ik})\Big]\eea where the internal sum is
taken over all word lists $\g_k \in \G$. Each term of the internal
sum is zero by the NNG's local detailed balance at fixed site $s_i$
because $P(\cdot,\cdot| X_0(\sN(s_i)))$ is the local transition
probability of the NNG conditioned on fixed neighborhood state
$X_0(\sN(s_i))$,
and the quotient $\pi'(G_{ik})$/$\pi'(G_0)$ has been shown to equal the
quotient $f_i(\g_k|X_0(\sN(s_i)))$/$f_i(X_0(s_i)|X_0(\sN(s_i)))$ where $f_i(\cdot|X_0(\sN(s_i)))$
are the values of the local specification at site $s_i$ of the NNGGS. In other words,
the NNGGS' local specification, by construction  %$f_i(\g_k|X_0(\sN(s_i)))$ --
(repeating the local transition step (2) in the NNG at fixed site
$s_i$, under the same neighborhood state $X_0(\sN(s_i))$), is the
stationary distribution of an auxillary Markov chain consisting of
varying local states at the single site $s_i$ and a fixed boundary
state at the sites in $\sN(s_i)$. We deduce that
$\Delta\pi'(G_0)=0$, i.e. the Gibbs distribution $\pi^{'}(G)$ of the
NNGGS is also the invariant distribution of the NNG.

\section{ Clique Potentials and community structure - analysis} The first example
concerns a 2-clique as in figure 1. With the local 0 state chosen to
be $Name=\{A,B\}$, and applying the above expression for
$F(x(E)|0(\sN(E)))$ in terms of the local specification which are in
turn calculated using the stationary distribution of the
corresponding local Markov chains (each with one site), we derived,
in the vanishing noise limit, the probabilities tabulated below.
This table gives a 2-clique potential for all possible
configurations with its neighborhood fixed at the local 0 state,
after using the natural symmetry in the problem. It shows this
2-clique has lowest energy when the two sites has the same single
name.
\begin{center}
\begin{figure}[ht]
  \includegraphics[width=0.28\textwidth]{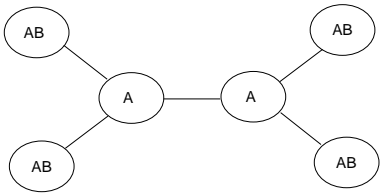}\\
  \caption{2-Clique}\label{figure:1}
\end{figure}
\end{center}

\begin{table}[h]
\begin{tabular*}{\hsize}{@{\extracolsep{\fill}}crr}\hline
$x(L)$ & $F(x(L)|0(\sN(L)))$ & $V(x(L))$ \\\hline A-AB & 1 & 0 \\ A-A & 2 & -0.6931
\\ A-B& 0.5 & 0.6931 \\\hline
\end{tabular*}
\end{table}
A more interesting example in the following graph has two 3-cliques
and a 2-clique that bridges them. Using the above procedure (and
labeling the sites starting from the site on the bridge, e.g. B-A-AB
means the site on the bridge has word list {B}), we calculate the
clique potential for the 3-clique and tabulate its values below.
\begin{center}
\begin{figure}[ht]
  \includegraphics[width=0.28\textwidth]{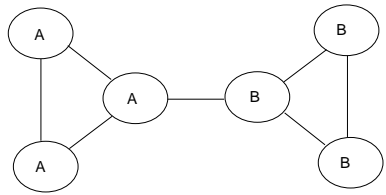}\\
  \caption{Example 2}\label{figure:2}
\end{figure}
\end{center}

\begin{table}[h]
\caption{Clique Potential for 3-Clique}
\begin{tabular*}{\hsize}{@{\extracolsep{\fill}}crr}\hline
$x(L)$ & $F(x(L)|0(\sN(L)))$ & $V(x(L))$ \\ \hline A-A-A & 15 &
-2.7080\\ AB-A-A & 3 & -1.0986 \\ B-A-A & 3/5 &
-0.5108
\\\hline
\end{tabular*}
\end{table}
In this way, we calculate the Gibbs potential for each network state
and show that multi-name states are ordered by their Gibbs energy
which we take to be a good measure of social tension in a particular
state. After the single-name ground state,
%where all nodes have the same single-name local state
the state in figure 2 has the second lowest energy. Significantly,
as the third main result and primary focus of this letter, the
naming or coloring that reveals the underlying cliques in the graph
also has the least energy amongst all multi-name states for given
community (clique) structure in a network. Moreover, it is a local
minimum, i.e. any one-step change of this configuration will
increase its energy: $H(AAA-BBB)=-4.7230$ and $H(AAA-ABB)=-2.8904$.
These lowest energy multi-name states are therefore the most likely
ones to be found by simulated annealing of the NNG. This is
consistent with the results on meta-stable states in \cite{Lu2009},
but have the advantage, in the NNG, of being persistent (recurrent),
thence, enhancing its community-detection capability over the
original NG.

\section{Stationary probability distributions of the NNG on small randomized networks - simulations}
We conclude with a brief discussion of the method used to compute the invariant probability distributions of the NNG on a family of small randomized networks consisting of 60 nodes - these network graphs are computer generated.  Using a total word list of cardinality two, we show that the NNG efficiently finds the 40:20 splits into community structures where two of the communities or subgraphs (by design of about 20 nodes each) have one name / color and the remaining subgraph of 20 nodes have the second name / color In addition, it also produces an invariant measure which, after discounting highest values achieved naturally at the total consensus or single-name states at the 0 and 60-nodes peaks, gives relative weights on the meta-stable states that provide information on community structure. For example, we refer to the symmetric 20 and 40 - nodes peaks in the following figures. About 100 million steps of the NNG was used to produce the invariant measure depicted.

\begin{center}
\begin{figure}[ht]
  \includegraphics[width=0.35\textwidth]{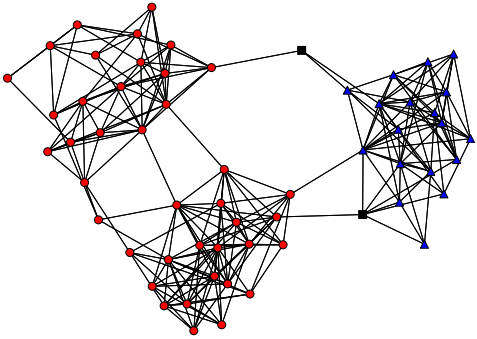}\\
  \caption{Blue triangle for A, red circle for B and black square for AB}\label{figure:4}
\end{figure}
\end{center}
\begin{center}
\begin{figure}[ht]
  \includegraphics[width=0.35\textwidth]{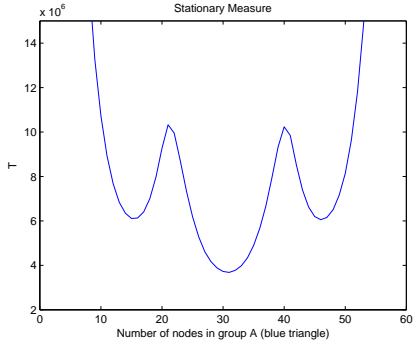}\\
  \caption{Invariant measure for NNG}\label{figure:3}
\end{figure}
\end{center}

\appendix[Gibbs potential from local specification]
%\section{Gibbs potential from local specification}
\subsection{Local specification}
In general, it's not easy to calculate the invariant measure of a local Markov chain, namely, its local specification $f_i(X(s_i)|X(\sN(s_i)))$.
However for small cliques and when there are only a small number of allowed words such as two in the three local states (A, B, AB) in the above examples, the transition in the Markov chain is totally decided by the message that
$s_i$ received from its neighbors.  Let $p_A(s_i)$,$p_B(s_i)$ be the probability for $s_i$ to receive  $A, B$ respectively. Their values can be calculated
from the neighborhood condition $X(\sN(s_i))$; more precisely the number of neighbors in state A, B and AB, are given by:
\bea \textstyle p_A(s_i)={\e\over 2} +(1-\e){\# \{s_j\in \sN(s_i)|X(s_j)=A\}+{1\over 2}\# \{s_j\in \sN(s_i)|X(s_j)=AB\}\over\# \{s_j\in \sN(s_i)\}}\\
\textstyle p_B(s_i)={\e\over 2} +(1-\e){\# \{s_j\in \sN(s_i)|X(s_j)=B\}+{1\over 2}\# \{s_j\in \sN(s_i)|X(s_j)=AB\}\over\# \{s_j\in \sN(s_i)\}}
\eea

Given the noise level $\e$ is arbitrarily small, we can take the limit $\e \go 0$:
$$\textstyle p_A(s_i)={\# \{s_j\in \sN(s_i)|X(s_j)=A\}+{1\over 2}\# \{s_j\in \sN(s_i)|X(s_j)=AB\}\over\# \{s_j\in \sN(s_i)\}}$$
$$\textstyle p_B(s_i)={\# \{s_j\in \sN(s_i)|X(s_j)=B\}+{1\over 2}\# \{s_j\in \sN(s_i)|X(s_j)=AB\}\over\# \{s_j\in \sN(s_i)\}}$$
Leaving aside the transitions involving a single-name state,
%$p_A$ or $p_B$ would be 0, and all following calculation using $\ln(p_A)$ and $\ln(p_A)$
%would fail.
that is, as long as one only considers the transition between two multi-name states, taking this limit is valid.
From detailed balance we get
$$\textstyle f_i(A|x(\sN(s_i))) p_B(s_i)=f_i(AB|x(\sN(s_i))) p_A(s_i)$$
$$\textstyle f_i(AB|x(\sN(s_i))) p_B(s_i)=f_i(B|x(\sN(s_i))) p_A(s_i)$$
from which we find the local specification:
$$\textstyle f_i(A|x(\sN(s_i)))=p_A^2(s_i)/Z_l(s_i)$$
$$\textstyle f_i(B|x(\sN(s_i)))=p_B^2(s_i)/Z_l(s_i)$$
$$\textstyle f_i(AB|x(\sN(s_i)))=p_A(s_i) p_B(s_i)/Z_l(s_i)$$
$$\textstyle Z_l(s_i)=p_A^2(s_i)+p_B^2(s_i)+p_A(s_i) p_B(s_i)$$

\subsection{Properties of  $F(x(E)|0(\sN(E)))$}
By definition,\\
 $ F(x(E)|0(\sN(E)))$ is given by:
$$\textstyle \prod_{i=1}^M {f_i(x(s_i)|X(s_1,...,s_{i-1}),0(s_{i+1},...,s_M),0(\sN(E)))\over f_i(0(s_i)|X(s_1,...,s_{i-1}),0(s_{i+1},...,s_M),0(\sN(E)))}$$
Several facts will help us to simplify this formula.
\benum
\item
In each factor of the above product, the conditionals in the numerator and denominator are the same. So if $x(s_i)=0(s_i)$, the factor is 1.
Therefore only the sites not in 0 state counts in the product. A direct consequence of this fact is $F(0(E)|0(\sN(E)))=1$.\\
\item
For any 1-Clique in this example:\\
$F(AB(s_i)|0(\sN(s_i)))=1$ according to the last paragraph, and
$$F(A(s_i)|0(\sN(s_i)))={f_i(A|0(\sN(E)))\over f_i(AB|0(\sN(E)))}={p_A(s_i)\over p_B(s_i)}=1.$$
Similarly $F(B(s_i)|0(\sN(s_i)))=1$, thus, the function $F$ on any 1-Clique has 0 value.
\item
Consider the clique $E=\{s_1,...,s_M\}$ and its two distinguished configurations which only differ at one site $s_m$.  By relabeling the sites we can write the
two configurations as $X(E)=\{x(s_1),...,x(s_{m-1}),x(s_m),0(s_{m+1}),...,0(s_M)\}$ and $Y(E)=\{x(s_1),...,x(s_{m-1}),y(s_m),0(s_{m+1}),...,0(s_M)\}$. In the ratio $F(Y(E)|0(\sN(E)))/F(X(E)|0(\sN(E)))$, the factors for 0 states do not appear and the factors for $s_1,...,s_{m-1}$ cancel. So we have the recursive relationship:
$$\textstyle  {F(Y(E)|0(\sN(E)))\over F(X(E)|0(\sN(E)))}={f_m(y(s_m)|X(s_1,...,s_{m-1}),0(s_{m+1},...,s_M),0(\sN(E)))\over f_m(x(s_m)|X(s_1,...,s_{m-1}),0(s_{m+1},...,s_M),0(\sN(E)))}.$$
Using this relationship we calculate the function value of $F$ recursively.
\eenum
\subsection{2-Clique}
In the example above, the network have one 2-Clique and two similar 3-Cliques. We need to calculate the function value of
$F$ for certain configurations on the given Cliques. In the following calculations, $p_A(s_i)$ and $p_B(s_i)$ are always calculated
according to the current neighbor configuration.
For the 2-Clique shown in Figure 1, we have
\bea &&F(AB-AB|0(\sN(E)))\\
&=&{F(0(E)|0(\sN(E)))}=1\\
&& F(A-AB|0(\sN(E)))\\
&=&\textstyle {f_1(A|0(s_2),0(\sN(E)))\over f_1(AB|0(s_2),0(\sN(E)))}\\
&=&\textstyle {p_A(s_1)\over p_B(s_1)}=1\\
 && F(A-A|0(\sN(E)))\\
 &=&\textstyle F(A-AB|0(\sN(E))){f_2(A|A(s_1),0(\sN(E)))\over f_2(AB|A(s_1),0(\sN(E)))}\\
  &=&\textstyle {p_A(s_2)\over p_B(s_2)}=2\\
&& F(A-B|0(\sN(E)))\\
&=&\textstyle F(A-AB|0(\sN(E))){f_2(B|A(s_1),0(\sN(E)))\over f_2(AB|A(s_1),0(\sN(E)))}\\
&=&\textstyle {p_B(s_2)\over p_A(s_2)}={1\over 2}.\eea

\subsection{3-Clique}
For the 3-Clique on the left in Figure 2,
%\begin{center}
%\begin{figure}[!htbp]
%  \includegraphics[width=0.4\textwidth]{NG3.eps}\\
%  \caption{3-Clique}\label{figure:3}
%\end{figure}
%\end{center}
\bea\textstyle && F(AB-AB-AB|0(\sN(E))) \textstyle ={F(0(E)|0(\sN(E)))}=1\\
\textstyle   &&F(AB-AB-A)|0(\sN(E)))\\
 &=&\textstyle{f_3(A|0(s_1,s_2),0(\sN(E)))\over f_3(AB|0(s_1,s_2),0(\sN(E)))} =\textstyle{p_A(s_3)\over p_B(s_3)}=1\\
 &&F(AB-A-A)|0(\sN(E)))\\
&=&\textstyle F(AB-AB-A|0(\sN(E))){f_2(A|0(s_1),A(s_3),0(\sN(E)))\over f_2(AB|0(s_1),A(s_3),0(\sN(E)))}\\
&=&\textstyle {p_A(s_2)\over p_B(s_2)}=3\\
 &&F(A-A-A)|0(\sN(E)))\\
 &=&\textstyle F(AB-A-A|0(\sN(E))){f_1(A|A(s_2,s_3),0(\sN(E)))\over f_1(AB|A(s_2,s_3),0(\sN(E)))}\\
&=&\textstyle 3{p_A(s_1)\over p_B(s_1)}=15\\
  &&F(B-A-A)|0(\sN(E)))\\
  &=& \textstyle F(AB-A-A|0(\sN(E))){f_1(B|A(s_2,s_3),0(\sN(E)))\over f_1(AB|A(s_2,s_3),0(\sN(E)))}\\
   &=&\textstyle 3{p_B(s_1)\over p_A(s_1)}={3\over 5}
\eea
From A-B symmetry and graph symmetry, we can get other function values from these results.

\subsection{Gibbs Potential}
The 3-Clique in Figure 2 has three 2-Cliques and three 1-Cliques embedded in it. Adding them, calling the sum "net clique potential" for the 3-Clique "L",
and using the fact that the terms for 2-Cliques cancel, the result turns out to be:
$V_{net}(L)=-\ln(F(L|\sN(L)))$.\\
Finally, the total Gibbs potential includes the clique potential of the 2-Clique and the net clique potential of two 3-Cliques,
\bea && H(AAA-BBB)\\
&=& V(A-B)+V_{net}(A-A-A)+V_{net}(B-B-B)\\
&=& -\{\ln(F(A-B|0(\sN(E))))+2\ln(F(A-A-A|0(\sN(E)))\}\\
&=& -4.7230
\eea
\bea    &&H(AAA-ABB)\\
&=& V(A-A)+V_{net}(A-A-A)+V_{net}(A-B-B)\\
&=& -\{\ln(F(A-A|0(\sN(E))))+\ln(F(A-A-A|0(\sN(E)))\\
&&+\ln(F(A-B-B|0(\sN(E)))\}\\
&=& -2.8904
\eea
%%%%%%%%%%%%%%%%%%%%%%%%%%%%%%%%%%%%%%%%%%%%%%%%%%%%

\textbf{Acknowledgments:} Research was sponsored by the Army Research Laboratory and was accomplished under Cooperative Agreement Number W911NF-09-2-0053 to the Social Cognitive Networks Acad Res Center at RPI.
%%%%%%%%%%%%%%%%%%%%%%%%%%%%%%%%%%%%%%%%%%%%%%%%%%%%%%%%%%%%%%%%%%%%%%%%%%%%%%

%%%%%%%%%%%%%%%%%%%%%%%%%%%%%%%%%%%%%%%%%%%%%%%%%%%%%%%%%%%%%%%%%%%%%%%%%%%%%%%
%%%%%%%%%%%%%%%%%%%%%%%%%%%%%%%%%%%%%%%%%%%%%%%%%%%%%%%%%%%%%%%%%%%%%%%%%%%%%%%

\begin{thebibliography}{}
\bibitem{Barabasi}R. Albert and A-L. Barabasi, Statistical Mechanics
of Complex Networks, \emph{Rev. Mod. Phys.}, 74, 47-97, 2002.
\bibitem{AlbertJeongBarabasiNature} R. Albert H. Jeong and A-L. Barabasi \emph{Nature}, 401, 130, 1999.
\bibitem{WattsStrogatzNature} D. Watts and S. Strogatz, \emph{Nature},
393, 440, 1998.
\bibitem{Angluin1980}D. Angluin. In \emph{Proceedings of the 12th ACM Symposium on Theory of Computing}, page
82-93, New York, 1980. ACM.
%\bibitem{Baronchelli2006}
%A. Baronchelli, L. DallAsta, A. Barrat, and V. Loreto. \emph{Phys.
%Rev. E}, {73}(015102(R)), 2006.
\bibitem{Baronchelli2006-2}
A. Baronchelli, M. Felici, E. Caglioti, V. Loreto, and L. Steels
\emph{J. Stat. Mech.: Theory Exp.}, (P06014), 2006
\bibitem{Ben-Naim1996}E. Ben-Naim, L. Frachebourg, and P. L. Krapivsky. \emph{Phys. Rev. E}, 53(3078), 1996.
\bibitem{Bremaud}P. Bremaud. Markov chains: Gibbs fields, Monte Carlo simulation, and queues. Springer,
1999.
\bibitem{Casella1992}G. Casella and E. I. George. \emph{The American Stat.}, 46(3), 1992.
%\bibitem{Castellano2003}C. Castellano and A. Vespignani D.
%Vilone. \emph{Europhys. Lett.}, 63(153), 2003
\bibitem{Castellano2005}C. Castellano, V. Loreto, A. Barrat, F. Cecconi, and D. Parisi. \emph{Phys. Rev. E,} 71(066107),
2005.
\bibitem{Collier2004}T. C. Collier and C. E. Taylor. J. \emph{Parallel Distrib. Comput.}, 64(866), 2004.
%\bibitem{Dall2002}J. Dall and M. Christensen. \emph{Phys. Rev. E}, 66(016121), 2002.
%\bibitem{Dall'Asta2006}L. Dall'Asta, A. Baronchelli, A. Barrat, and V. Loreto. \emph{Europhys. Lett.}, 73(969), 2006.
\bibitem{Dall'Asta2006}L. Dall'Asta, A. Baronchelli, A. Barrat, and V. Loreto. \emph{Phys. Rev. E}, 74(036105), 2006.
%\bibitem{Glazier1990}J. A. Glazier, M. P. Anderson, and G. S. Grest. \emph{Philos. Mag. B}, 62(615), 1990.
\bibitem{Grimmett1973}G. R. Grimmett.\emph{ Bull. London Math. Soc.}, 5, 1973.
%\bibitem{Hirschberg1980}D. S. Hirschberg and J. B. Sinclair. \emph{Commun. ACM}, 23(627), 1980.
\bibitem{Kaski1985}K. K. Kaski, J. Nieminen, and J. D. Gunton. \emph{Phys. Rev. B}, 31(2998), 1985.
\bibitem{Kumar1987}S. Kumar, J. D. Gunton, and K. K. Kaski. \emph{Phys. Rev. B}, 35(8517), 1987.
%\bibitem{Lee2005}Y. Lee, T. C. Collier, C. E. Taylor, and E. E. Stabler. In \emph{Proceedings of the Tenth
%International Symposium on Artificial Life and Robotics}, Beppu,
%Japan, 2005.
%\bibitem{LeLann1977}G. LeLann. In \emph{IFIP Congress Proceedings}, pages 155-160, Amsterdam, 1977. North
%Holland.
\bibitem{Limnebus} C. Lim and J. Nebus, \emph{Vorticity, Stat Mech and Monte-Carlo simulations}, Springer-Verlag 2006.
\bibitem{Lucomm} Q. Lu, G. Korniss, and B.K. Szymanski, The Naming Game in social networks: community formation and consensus
engineering, \emph{J. Economic Interaction and Coordination}, 4,
221-235 (2009).
\bibitem{Lu2008} Q. Lu, G. Korniss and, and B. K. Szymanski. \emph{Phys. Rev. E}, 77(016111), 2008.
%\bibitem{Lu2006}Q. Lu, G. Korniss, and B. K. Szymanski. In \emph{Proceedings of the 2006 American Association
%for Arti¡¥cial Intelligence Fall Symposium Series, Interaction and
%Emergent Phenomena in Societies of Agents}, page 148-155, Menlo
%Park, CA, 2006. AAAI Press.
\bibitem{Lu2009}Qiming Lu. PhD thesis, R.P.I., NY, 2009.
%\bibitem{Malpani2000}N. Malpani, J. Welch, , and N. Vaidya. In \emph{Proceedings of the Fourth International
%Workshop on Discrete Algorithms and Methods for Mobile Computing and
%Communications}, page 96-103, New York, 2000. ACM.
\bibitem{Matsen2004}F.A. Matsen and M.A. Nowak. \emph{Proc. Natl. Acad. Sci. U.S.A.}, 101(18053), 2004.
\bibitem{Meester1996}R. Meester and R. Roy. \emph{Continuum Percolation}. Cambridge University Press, Cambridge,
England, 1996.


\bibitem{Penrose2003}M. Penrose, \emph{Random Geometric Graphs}. Oxford University Press, New York,
2003.
\bibitem{Nowak2001}M.A. Nowak and N.L. Komarova. \emph{Science}, 291(114), 2001.
%\bibitem{Nowak1999}M.A. Nowak, J.B. Plotkin, and C. Krakauer. J. \emph{Theor. Biol}, 200(147), 1999.
\bibitem{Roland1990}C. Roland and M. Grant. \emph{Phys. Rev. B}, 41(4663), 1990.
\bibitem{Stanley} H.E. Stanley, \emph{Intro to Phase  transitions and Critical
phenomena}, Oxford U. Press, 1982.
\bibitem{LeeAllisonAbbottStanley} Y. Li, A. Allison, D. Abbott, H.E. Stanley, \emph{PRL}, 91(22), 220601-1, 2003.

\bibitem{Motter06} C. Zhou, A.E. Motter and J. Kurths \emph{Phys. Rev. Lett.}, 96, 034101, 2006.
%\bibitem{Motte10} A.E. Motter,  arXiv:1003.2465, 2010.
%\bibitem{Martens10} A.E. Martens, C.R. Laing and S.H. Strogatz,  arXiv:0910.5389v2, 2010.
%\bibitem{Huang2006} L. Huang, K. Park, Y.C. Lai, L. Yang and K. Yang \emph{Phys. Rev. Lett.}, 97,
164101, 2006.
\bibitem{Arenas2006} A. Arenas, A. Diaz-Guilera, C.J. Perez-Vicente \emph{Phys. Rev. Lett.}, 96,
114102, 2006.
\bibitem{Chavez2005} M. Chavez, D.U. Hwang, A. Amann, H.G.E. Hentschel and S. Boccaletti \emph{Phys. Rev. Lett.},  94, 218701 ,2005.
\bibitem{StanleyPNAS2000} L. Amaral, A. Scala, M. Barthelemy and H.E.
Stanley, \emph{Proc. Natl. Acad. Sci. U.S.A.}, 97(21), 11149, 2000.
\bibitem{Earn2006} D. J. D. Earn and S. A. Levin, \emph{Proc. Natl. Acad. Sci. U.S.A.}, 103 , 11, 2006.
%\bibitem{Hopfield2001} J. J. Hopfield and C. D. Brody \emph{Proc. Natl. Acad. Sci. U.S.A.}, 98,3  , 2001.
\bibitem{Girwan2002} M. Girvan and M. E. J. Newman \emph{Proc. Natl. Acad. Sci. U.S.A.} 99, 12, 2002.
\bibitem{NewmanPNAS2006} M. E. J. Newman \emph{Proc. Natl. Acad. Sci. U.S.A.} 103, 8577, 2006.
\bibitem{Chauhan2009} S.Chauhan, M.Girvan and E. Ott,  arXiv:0911.2735v1, 2009.
\bibitem{Thattai} M. Thattai and A. Oudenaarden \emph{Proc. Natl. Acad. Sci. U.S.A.} 98,15,2001.
\bibitem{Nishimoribook} H. Nishimori, \emph{Stat Phys of Spin-Glass and Info processing : an Intro.}, Oxford U. Press.


\end{thebibliography}
\end {document}